\documentclass[twocolumn,notitlepage,prl,superscriptaddress,showpacs]{revtex4-1}
\usepackage{hyperref}
\usepackage[usenames,dvipsnames]{color}
\usepackage{amsmath}
\usepackage[english]{babel}
\usepackage{amssymb}
\usepackage{graphicx}
\usepackage{epstopdf}
\usepackage{tabularx}
\usepackage{multirow}
\usepackage[]{todonotes}

\begin{document}
\title{Algorithmic Matsubara Integration for Hubbard-like models}

\author{Amir Taheridehkordi}
\affiliation{Department of Physics and Physical Oceanography, Memorial University of Newfoundland, St. John's, Newfoundland \& Labrador A1B 3X7, Canada} 
\author{S. H. Curnoe}
\affiliation{Department of Physics and Physical Oceanography, Memorial University of Newfoundland, St. John's, Newfoundland \& Labrador A1B 3X7, Canada} 
\author{J. P. F. LeBlanc}
\email{jleblanc@mun.ca}
\affiliation{Department of Physics and Physical Oceanography, Memorial University of Newfoundland, St. John's, Newfoundland \& Labrador A1B 3X7, Canada} 

\date{\today}

\begin{abstract}
We present an algorithm to evaluate Matsubara sums for Feynman diagrams comprised of  bare Green's functions with single-band dispersions with local $U$ Hubbard interaction vertices. The algorithm provides an exact construction of the analytic result for the frequency integrals of a diagram that can then be evaluated for all parameters $U$, temperature $T$, chemical potential $\mu$, external frequencies and internal/external momenta. This method allows for symbolic analytic continuation of results to the real frequency axis, avoiding any ill-posed numerical procedure. When combined with diagrammatic Monte-Carlo, this method can be used to simultaneously evaluate diagrams throughout the entire $T$-$U$-$\mu$ phase space of Hubbard-like models at minimal computational expense.
\end{abstract}

\maketitle

The Hubbard model\cite{benchmarks} is a cornerstone of correlated electron physics and plays an important role as a testbed for the development of numerical algorithms.
Among modern numerical tools, Diagrammatic Monte Carlo (DiagMC) is a powerful technique which performs integrals arising from perturbative expansions by sampling classes of connected Feynman diagrams.\cite{vanhoucke,vanhoucke:natphys,kozik:2010,rossi2017determinant}
Other algorithms have been developed from expansions around non-perturbative dynamical mean-field theory\cite{toschi:2007,rubtsov:2008,rubtsov:2009}  as well as so-called `bold' extensions to DiagMC with a variety possible of resummation schemes.\cite{nikolay:bold,kulagin:2013}  However, it was recently shown\cite{kozik2015nonexistence,tarantino:2017} that the resummation of the skeleton Feynman diagrammatic series for systems with the Hubbard interaction will lead to a false convergence towards an unphysical branch, due to the Riemann series theorem at strong interactions, while the series based on bare Green's functions always converges to the expected physical result.\cite{kozik2015nonexistence} As a result, expressing the perturbation series in terms of bare Green's functions (and bare vertices) might be preferable.

In the case of Hubbard-like  models,\cite{benchmarks} since each bare vertex is unstructured ($U$) in principal one needs only to compute the series of integrals over internal spatial (momentum) and time (frequency, commonly computed as a sum over Matsubara frequencies) variables for each diagram. Despite the conceptual simplicity of this proposal, in practice the problem remains a challenge.  One difficulty lies in the factorial scaling of the number of diagrams one must sample as the interaction order increases.\cite{rossi2017determinant,simkovic2017determinant,ferrero:2018} 
Another is the poor convergence of sums over Matsubara frequencies, since the set of Matsubara frequencies [$i\omega_n=i\frac{\pi}{\beta}(2n+1)\ \ \text{or} \ \ i\frac{\pi}{\beta}(2n)$ for fermions and bosons respectively] compresses as the temperature $T=1/\beta$ decreases.
 Worse still is that numerical results by necessity express external lines of the Feynman diagrams in terms of Matsubara frequencies.   The numerical process of analytic continuation of Matsubara frequencies to real frequencies is ill-posed, and while procedures such as maximum entropy inversion or pad\'e approximants have become standard and codes to implement these procedures are widely available,\cite{bergeron:2016,Levy2016,ALPS20,Gaenko17} the problem of analytic continuation remains a roadblock to providing reliable theoretical results to correlated many-body problems.

In this letter we propose a method which we call Algorithmic Matsubara Integration (AMI) in which we utilize the residue theorem to compute summations over independent Matsubara frequencies. The result of the algorithm is an analytic expression for the temporal integrals of a diagram of arbitrary order in terms of internal and external momenta and external Matsubara frequencies, upon which one can impose a true analytic continuation $i\omega_n \to \omega +i0^+$.
We demonstrate the utility of our method by evaluating a variety of diagrams for the 2D Hubbard model on a square lattice and comment on the scaling of computational cost with complexity of the integrand (\emph{i.e.} expansion order).
\newline \indent \emph{Algorithm:} 
DiagMC typically samples the entire space of diagram topologies as well as sampling over internal variables such as a set of momenta $\{k_n\}=k_1,k_2,...,k_n$ and a set of frequencies $\{\nu_n\}=\nu_1,\nu_2,...,\nu_n$.\cite{vanhoucke} 
Our aim is to reduce the space of sampling for DiagMC from $\{k_n,\nu_n\}\to\{k_n\}$ by algorithmic evaluation of the analytic result of the $\{\nu_n\}$ integrals. By evaluating the sums over Matsubara frequencies algorithmically we completely remove the need to probe the frequency (time) configuration space.  What remains for DiagMC is to traverse the space of diagram topologies and $\{k_n\}$ and use AMI to evaluate the full set of frequency integrals for each configuration. 

Making no assumptions about the topology of the diagram,
the general form of a diagram can be written as
\begin{align}
&\frac{U^{n_v}}{\beta^n}\sum\limits_{\{k_n\}} \sum \limits _{\{\nu_n\}}\prod
\limits_{j=1}^N G^j(\epsilon ^j, X^j ) = U^{n_v}\sum\limits_{\{k_n\}} I^{(n)}, \\
& I^{(n)}=\frac{1}{\beta^n}\sum \limits _{\{\nu_n\}}\prod
\limits_{j=1}^N G^j(\epsilon ^j, X^j ), \label{eqn:goal}
\end{align}
where $n_v$ is the order (the number of vertices) of the diagram, $n$ 
is the number of summations over Matsubara frequencies $\{\nu_n\}$  and 
internal momenta $\{k_n\}$, and $N$
is the number of internal lines representing bare Green's functions $G(\epsilon,X)$.
The bare Green's function of the $j$th internal line is
\begin{eqnarray}\label{E: Green_Function}
G^j(\epsilon^j, X^j) =  \frac{1}{ X^j - \epsilon^j},
\end{eqnarray}  
where $X^j$ is the frequency and $\epsilon^j = \epsilon^j(k_j)$ is the free particle dispersion. 
Constraints derived from energy and momentum conservation at each vertex 
allow us to express these quantities as linear combinations of internal $\{\nu_n, {k}_n\}$ and external $\{ \nu_\gamma, k_\gamma\}$ frequencies and momenta, where $k_j  =  \sum_{\ell=1}^m \alpha_{\ell}^j k_{\ell} $ and $X^j   =   \sum_{\ell=1}^{m}i\alpha_\ell^j \nu_{\ell}$, where 
%
 $\gamma = m-n$ is the number of unconstrained external 
frequencies.
The coefficients $\alpha_\ell^j$ are numbers which have only three possible values:  zero, plus one or 
minus one. 
This allows us to 
 represent $G^j$ as an array of length $m+1$ of the form
\begin{eqnarray}\label{E: Green_Function_Array}
G^j(X^j) \to [\epsilon^j, \vec \alpha^j],
\end{eqnarray} 
where $\vec \alpha^j = (\alpha_1^{j}, ..., \alpha_{m}^{j})$. 
Given our array representation of each $G^j$, we construct a nested array to represent the product of $G^j$ which appears in Eq.~(\ref{eqn:goal}),
\begin{eqnarray}\label{E: Only_One_New_General_Sum_3}
\prod_{j=1} ^ N G^j(\epsilon^j, X^j) \to \bigg [[\epsilon^1, \vec \alpha^1]; [\epsilon^2, \vec \alpha^2]; ...; [\epsilon^N, \vec \alpha^N] \bigg].
\end{eqnarray}
The size of this array is $N\times(m+1)$.
As we shall show, this representation 
carries all the information we need to compute the summations in Eq.~(\ref{eqn:goal}).

To begin the algorithm, we subdivide the original problem to the summation over a single frequency $\nu_p$, and the remaining frequencies $\nu_n\neq \nu_p$,
\begin{align}
&I^{(n)}=\sum_{ \{\nu_n\}, \nu_n\neq \nu_p } I_p, \label{eqn:onesum} \\
&I_p = \sum_{\nu_p} \prod_{j=1} ^ N G^j(\epsilon^j, X^j_m).\label{E:Only_One_New_General_Sum}
\end{align}  

Central to computing Eq.~(\ref{E:Only_One_New_General_Sum}) is the identification of the set of simple poles of the Green's functions.  
The pole of the  $j$th Green's function with respect to the frequency $\nu_p$ 
exists so long as the coefficient $\alpha_p^j$  is non-zero, and is given by 
\begin{eqnarray}\label{E: Poles_one_h(z)}
z_p ^{(j)} = -\alpha_{p} ^{j} ( -\epsilon^j + \sum_{\ell=1, \ell\neq p} ^m i\alpha_\ell^{j}\nu_\ell)  \ \ \ \ \text{for} \ \ \ \alpha_{p}^{j} \neq 0 .
\end{eqnarray}
The number of simple poles for $\nu_p$ is
$r_p = \sum _{j=i}^{N} |\alpha_p^i|$, which occur in $r_p$ of $N$ total 
Green's functions in the product of Eq.~(\ref{E:Only_One_New_General_Sum}). 
We label these $r_p$ Green's functions as $G^{i_1}$, $G^{i_2}$, ..., $G^{i_{r_p}}$, and 
the set of simple poles will be denoted by $\{z_p^{(i_\ell)}\}_{\ell=1,2,...,r_p}$. 
Assuming all $z_p^{(i_\ell)}$ poles to be simple, the residue of each is
\begin{eqnarray}\label{E: Res_Th_Res_h(z)}
\alpha_p^{i_\ell}\prod_{j \neq i_\ell} G^j(\alpha_p^{j}z_p^{(i_\ell)} + \sum _{\ell \neq p} i\alpha_{\ell}^{j}\nu_{\ell}).
\end{eqnarray}
Note the sign $\alpha_p^{i_\ell}$ that is attached to this result.

To calculate the summation over the fermionic frequency $\nu_p$ in Eq.~(\ref{E:Only_One_New_General_Sum}) we use the residue theorem,
\begin{eqnarray}\label{E: Res_Th}
\sum_{\nu_p} h(i\nu_p) = \beta \sum_{z_p} f(z_p){\rm Res}[h(z)]_{z_p},
\end{eqnarray}
where $f(z)$ is the Fermi function
and $z_p$ are the poles of $h(z)$. 
%
%
Applying (\ref {E: Res_Th}) to the summation (\ref{E:Only_One_New_General_Sum}) and using (\ref {E: Res_Th_Res_h(z)}), we find the result:
\begin{widetext}
\begin{eqnarray}\label{eqn:result_recursion}
I_p = \alpha_p^{i_1}\beta f(z_p^{(i_1)}) \prod_{j \neq i_1} G^j(\alpha_p^{j}z_p^{(i_1)} + \sum _{{\ell} \neq p} i\alpha_{\ell}^{j}\nu_{\ell})+ 
\alpha_p^{i_2}\beta f(z_p^{(i_2)}) \prod_{j \neq i_2} G^j(\alpha_p^{j}z_p^{(i_2)} + \sum _{{\ell} \neq p} i\alpha_{\ell}^{j}\nu_{\ell}) \\ \nonumber + ... +
\alpha_p^{i_{r_p}}\beta f(z_{p}^{(i_{r_p})}) \prod_{j\neq i_{r_p}} G^j(\alpha_p^{j}z_{p}^{(i_{r_p})} + \sum _{{\ell} \neq p} i\alpha_{\ell}^{j}\nu_{\ell}). 
\end{eqnarray}  
\end{widetext}
The Fermi function is evaluated as
\begin{eqnarray}\label{E: Fermi_Simpl_diff}
f(z_p^{(i_\ell)}) =  \frac{1}{\sigma \exp(-\beta \alpha_p^{i_\ell}\epsilon^{i_\ell}) + 1},
\end{eqnarray}  
where  $\sigma$ is a sign given by
\begin{align}\label{E: Fermi_sign}
&\sigma(z_p^{i_\ell}) = \exp (i\beta\sum_{\ell\neq p} \alpha_\ell^{i_\ell} \nu_\ell),
\end{align}  
that is, $\sigma = -1$ if there are an odd number of fermionic frequencies in the sum over
$\ell$, otherwise $\sigma = 1$.
Therefore $f(z_p^{(i_\ell)})$ is independent of Matsubara frequencies and only depends on the real energy dispersion, though its character might switch from fermionic to bosonic. 

We have thus evaluated (\ref{E:Only_One_New_General_Sum}), a single
frequency summation.  There are $r_p$ terms in the result, and each term in this result contains a product of $N-1$ Green's functions, which may 
be represented as a $(N-1)\times (m+1)$ dimensional 
array in the form (\ref{E: Only_One_New_General_Sum_3}). 
These arrays may be arranged into a single nested array of
size $r_p \times (N-1) \times (m+1)$. 

We make use of this result to 
calculate {\em all} of the summations in  Eq.~(\ref{eqn:goal}) 
using a recursive procedure.
Without loss of generality we (arbitrarily) label the independent frequencies in the diagram as $\nu_1, \nu_2, \ldots \nu_n$, and perform the summations in this
order.
Each step of the procedure corresponds to the 
evaluation of one frequency summation. 
At the beginning of the procedure, the Feynman diagram that is to be evaluated is
represented as a $1\times N\times (m+1)$ dimensional array from which
the poles of $\nu_1$ are extracted. After the
first summation is computed using (\ref {eqn:result_recursion}), the result is stored in a $r_1\times (N-1) \times (m+1)$ 
dimensional array, and the poles of $\nu_2$ in this result are extracted. Subsequent steps will reduce the second dimension
by one on each step, but the first dimension will increase according to the 
number of poles. 
When all summations have been completed 
all that remains are residues defined by a set of $\alpha_j^p$ that are zero except for the $\gamma$ external frequencies.

To implement this procedure computationally we define
the following objects:
\begin{itemize}
\item the arrays $R_p$ representing the configurations of Green's functions after the $p$th summation (described above),

\item the sets of poles $P_p$ for $\nu_p$ in the configuration of Green's functions represented by $R_{p-1}$,

\item the set of signs $S_p$ of the residues
for each pole (the $\alpha_p^{i_\ell}$ in Eq.~(\ref {E: Res_Th_Res_h(z)})). 
\end{itemize}
The array of poles corresponding to $\nu_p$ has entries
\begin{eqnarray}\label{E: poles_array}
P_p = [P_p^{(1)}, P_p^{(2)}, ..., P_p^{(r_{(p-1)})}],
\end{eqnarray}  
with
\begin{eqnarray}\label{E: nu_p_poles_array}
P_p^{(\ell)} = [z_{p,\ell}^{(i_1)}, z_{p,\ell}^{(i_2)}, ..., z_{p,\ell}^{(i_{r_{\ell}} )}].
\end{eqnarray}  
We note that $P_p^{(\ell)}$ is the array of poles for $\nu_p$ in the residue of the $\ell$th pole for $\nu_{p-1}$ stored in the previous configuration of Green's functions, $R_{p-1}$. Similarly we have an array of signs with the same dimensions as $P_p$:
\begin{eqnarray}\label{E: sign_array}
S_p = [S_p^{(1)}, S_p^{(2)}, ..., S_p^{(r_{(p-1)})}],
\end{eqnarray}  
with
\begin{eqnarray}\label{E: nu_s_poles_array}
S_p^{(\ell)} = [\alpha_{p,\ell}^{(i_1)}, \alpha_{p,\ell}^{(i_2)}, ..., \alpha_{p,\ell}^{(i_{r_{\ell}} )}].
\end{eqnarray}  
where $\alpha_{p,\ell}$ are the nonzero coefficients of $\nu_p$ of the previous configuration of Green's functions, $R_{p-1}$.


Using these arrays, 
the full analytic result for Eq.~(\ref {eqn:goal}) is given by
\begin{eqnarray}\label{E: New_General_Sum_Again_Fianl_Again}
I^{(n)} = \frac{1}{\beta^n} \sum_{\{\nu_{n}\}} \prod_{j=1} ^ N G^j(\epsilon^j, X^j_m)  = K \cdot R_n,
\end{eqnarray}  
where
\begin{eqnarray}\label{E: Khorshid_Again}
K = (S_1 * f(P_1)) \times (S_2 * f(P_2)) \times ... \times (S_n * f(P_n)). \nonumber \\
\end{eqnarray}  
In this expression, $f(P_p)$ is the Fermi function of an array with elements
given by
\begin{eqnarray}\label{E: f_Op_new_compact}
[f(P_p)]_\ell^i = f(z_{p,\ell}^{(i)}),
\end{eqnarray}
and  the operations `$*$', `$\times$', and `$\cdot$' are defined by
\begin{eqnarray}\label{E: Dot_Gen_Compact}
(C * D)_i^j & = &  C_i^jD_i^j \equiv G_i^j, \nonumber \\ 
(G \times H)_i^j & =  & G_iH_i^j, \nonumber \\ 
H \cdot C  & = &  \sum_i H_i C_i. \nonumber 
\end{eqnarray}  
Equations (\ref{E: New_General_Sum_Again_Fianl_Again}) and (\ref{E: Khorshid_Again}) are obtained under the presumption that all of the poles are simple poles.  Poles with higher multiplicity are equivalent to multiple simple poles and therefore the result of Eq.~(\ref{E: New_General_Sum_Again_Fianl_Again}) holds even when poles with higher multiplicity arise.  However, it is not the ideal representation since upon evaluation one will find cancelling divergent terms which sum to non-zero values, causing numerical instability.  This problem can be avoided by generalizing for poles with multiplicity $M$. If $h(z)$ has a pole of order $M$ at $z=z_0$, then the residue is given by
\begin{eqnarray}\label{E: Res_mul}
Res[h(z_0)] = \frac {1} {(M-1)!} \lim_{z\to z_0} \frac {d^{M-1}}{dz^{M-1}} \bigg \{(z-z_0)^Mh(z)\bigg \}. \nonumber \\
\end{eqnarray}
In order to analytically evaluate arbitrary order derivatives, we employ the method of automatic differentiation which requires only knowledge of the first derivative and repeated application of chain rules. The first derivative with respect to $i\nu_p$ of the multiplication of $N$ Green's function is given via chain rule as
\begin{eqnarray}\label{E: Derivation_1_func}
\frac{d}{d(i\nu_p)}(\prod_{j=1} ^ N G^j(\epsilon^j, X^j_m)) = \sum_{i=1}^{N} \frac{dG^i}{d(i\nu_p)} \prod_{j\neq i} G^j(\epsilon^j, X^j_m). \nonumber \\
\end{eqnarray}  
The first derivative of one of the Green's function with respect to $i\nu_p$ in the array representation can then be performed by returning two Green's functions,
\begin{eqnarray}\label{E: Derivation_G_array}
\frac{dG^i(\epsilon^i, X_m^i)}{d(i\nu_p)} \to \bigg [ [\epsilon^i,X_m^i]; [-\alpha_p^i \epsilon^i, \alpha_p^i X_m^i] \bigg].
\end{eqnarray}  
The $(M-1)$th order derivative can be computed by iterating (\ref {E: Derivation_1_func}).  We therefore are able to express the residue for poles of $i\nu_p$ with any multiplicity using our symbolic representation. The only significant difference is that in the presence of multiple poles the entries of the $S$ array are $\pm \frac{1}{(M-1)!}$ instead of only $\pm1$.  The structures of $P$ and $R$ arrays remain the same but with additional terms arising from the chain rules, Eqs. (\ref{E: New_General_Sum_Again_Fianl_Again}) and (\ref{E: Khorshid_Again}) remain valid and are used to construct the final result.

We emphasize that since the result is symbolic in the set of yet-defined external frequencies $\{ \nu_\gamma \}$, at the final step one can replace $i\nu_\gamma \to \nu_\gamma + i0^+$ just as in a standard analytic continuation.  This eliminates the need for ill-posed numerical methods of analytic continuation in diagrammatics of Hubbard-like models. The method requires both the time to construct the solution, $t_c$, and the evaluation time, $t_e$, for each of $\gamma$ external variables.  We therefore expect the scaling will go as $\gamma t_{e}+ t_{c}$ where $t_{c}$ is typically larger than $t_{e}$.

\begin{figure}
\centering
\includegraphics[width=0.3\linewidth]{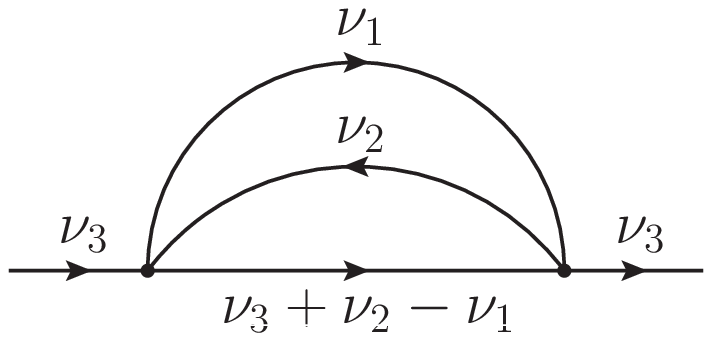}
\includegraphics[width=0.3\linewidth]{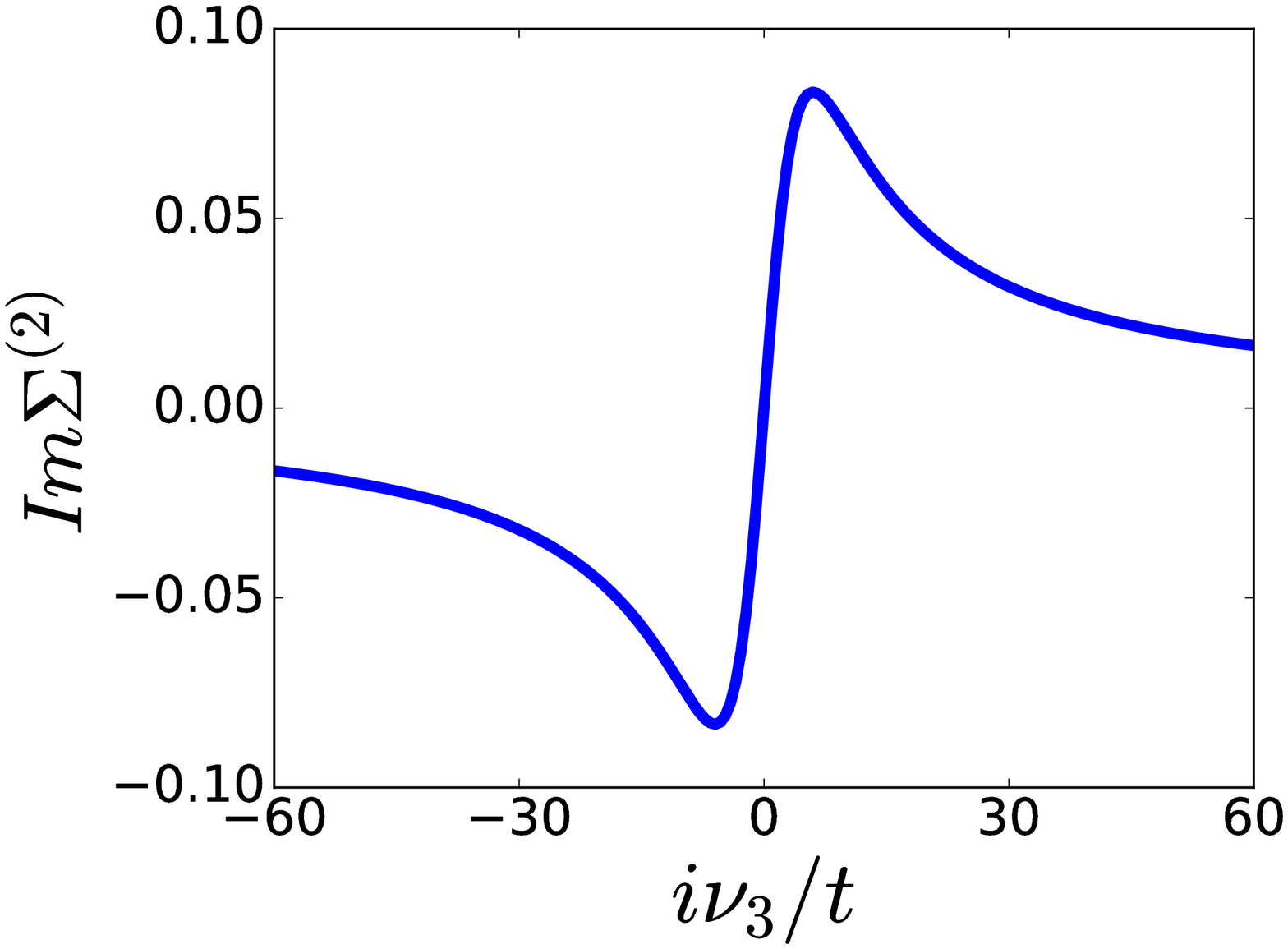}
\includegraphics[width=0.3\linewidth]{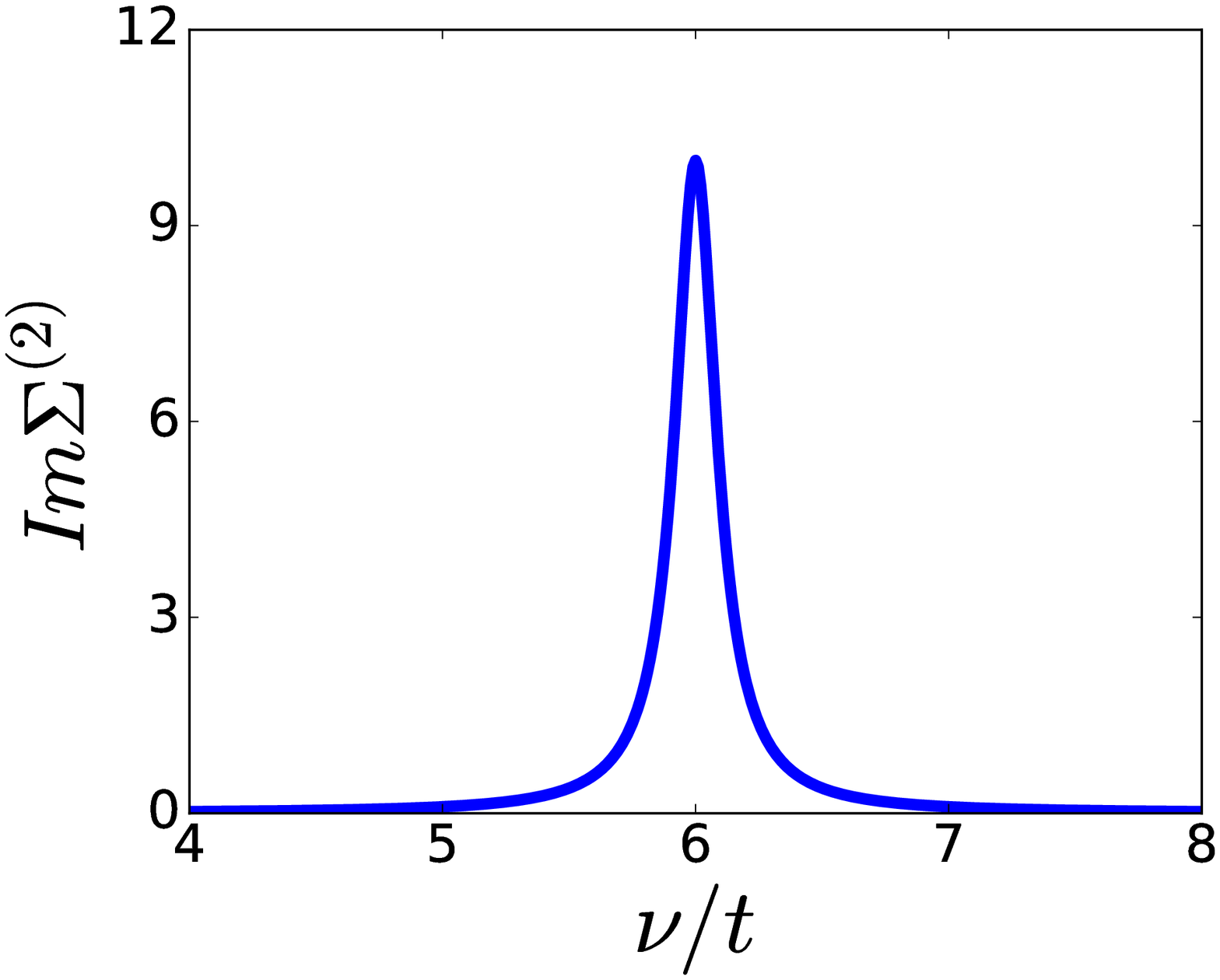}\\
\includegraphics[width=0.3\linewidth]{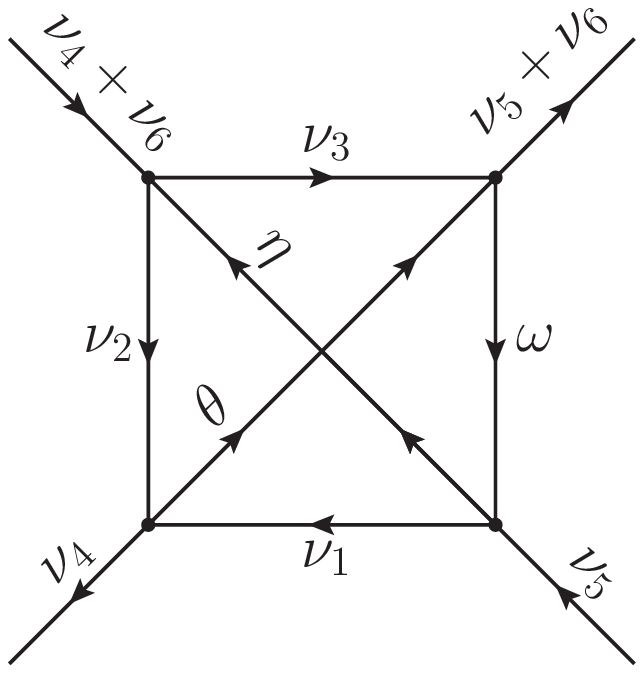}
\includegraphics[width=0.3\linewidth]{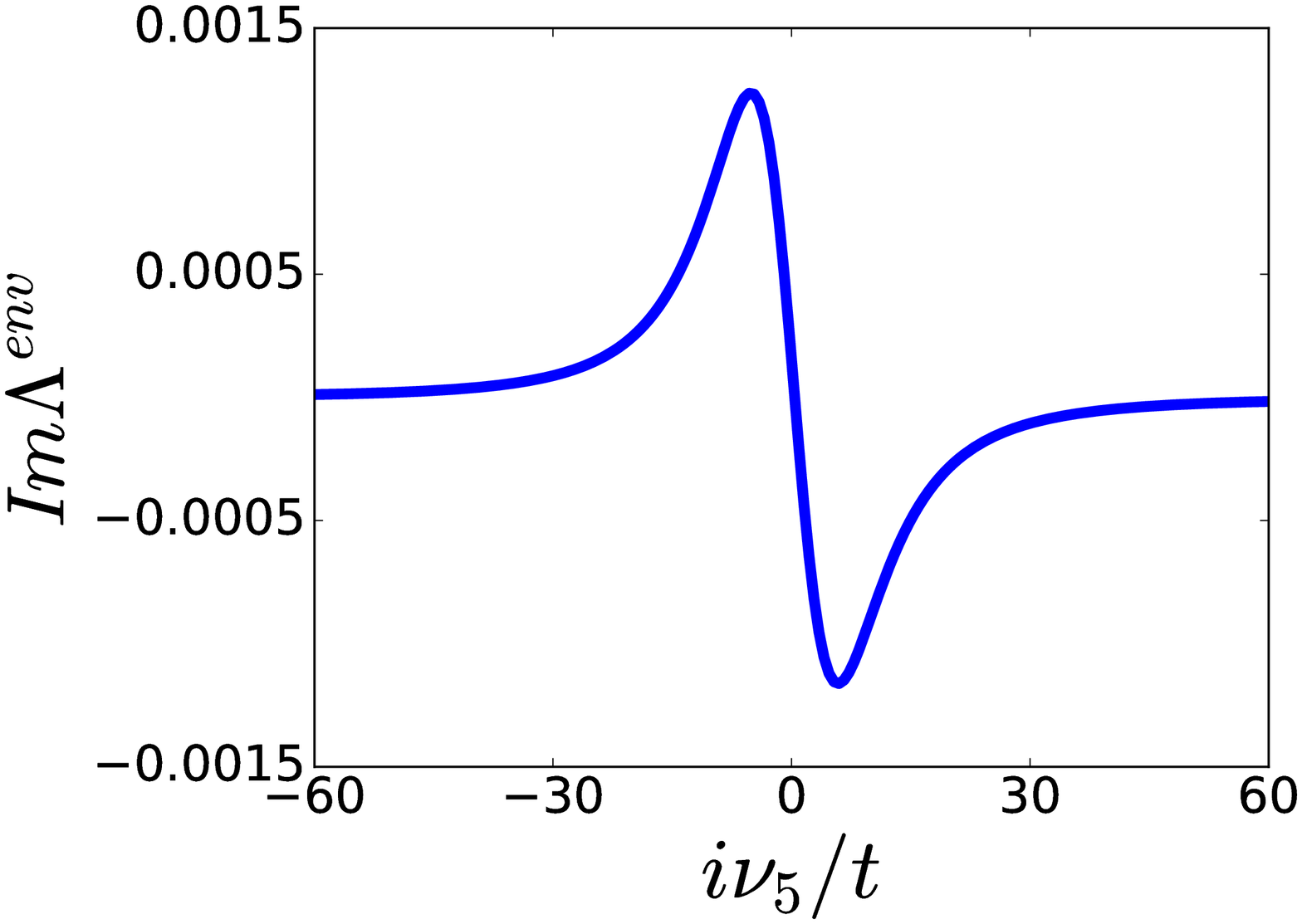}
\includegraphics[width=0.3\linewidth]{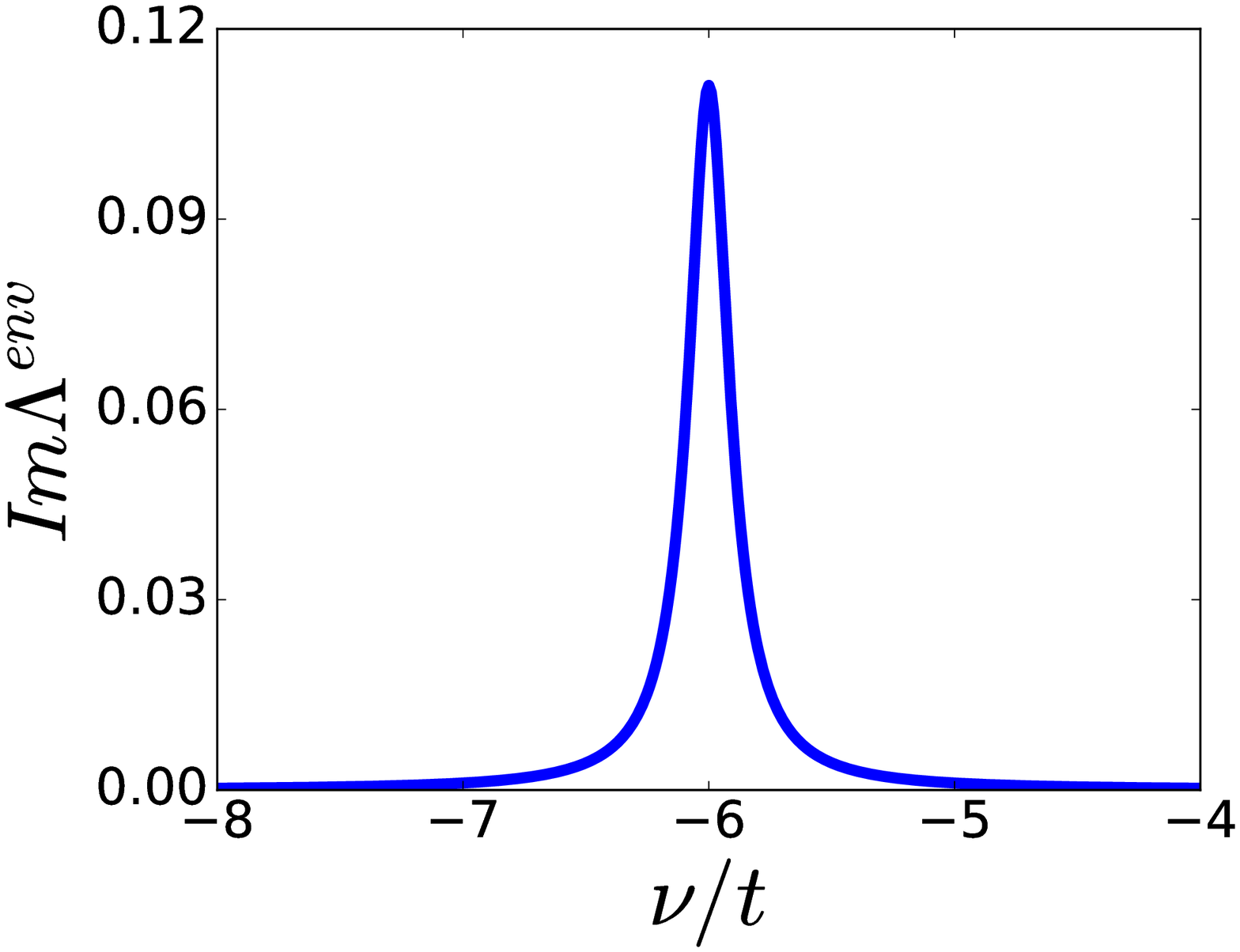} \\
\includegraphics[width=0.3\linewidth]{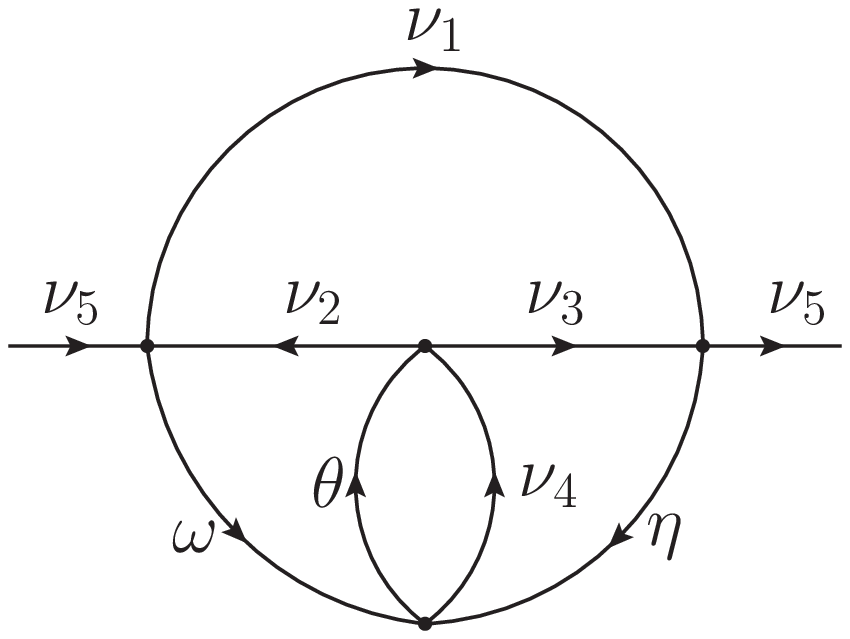}
\includegraphics[width=0.3\linewidth]{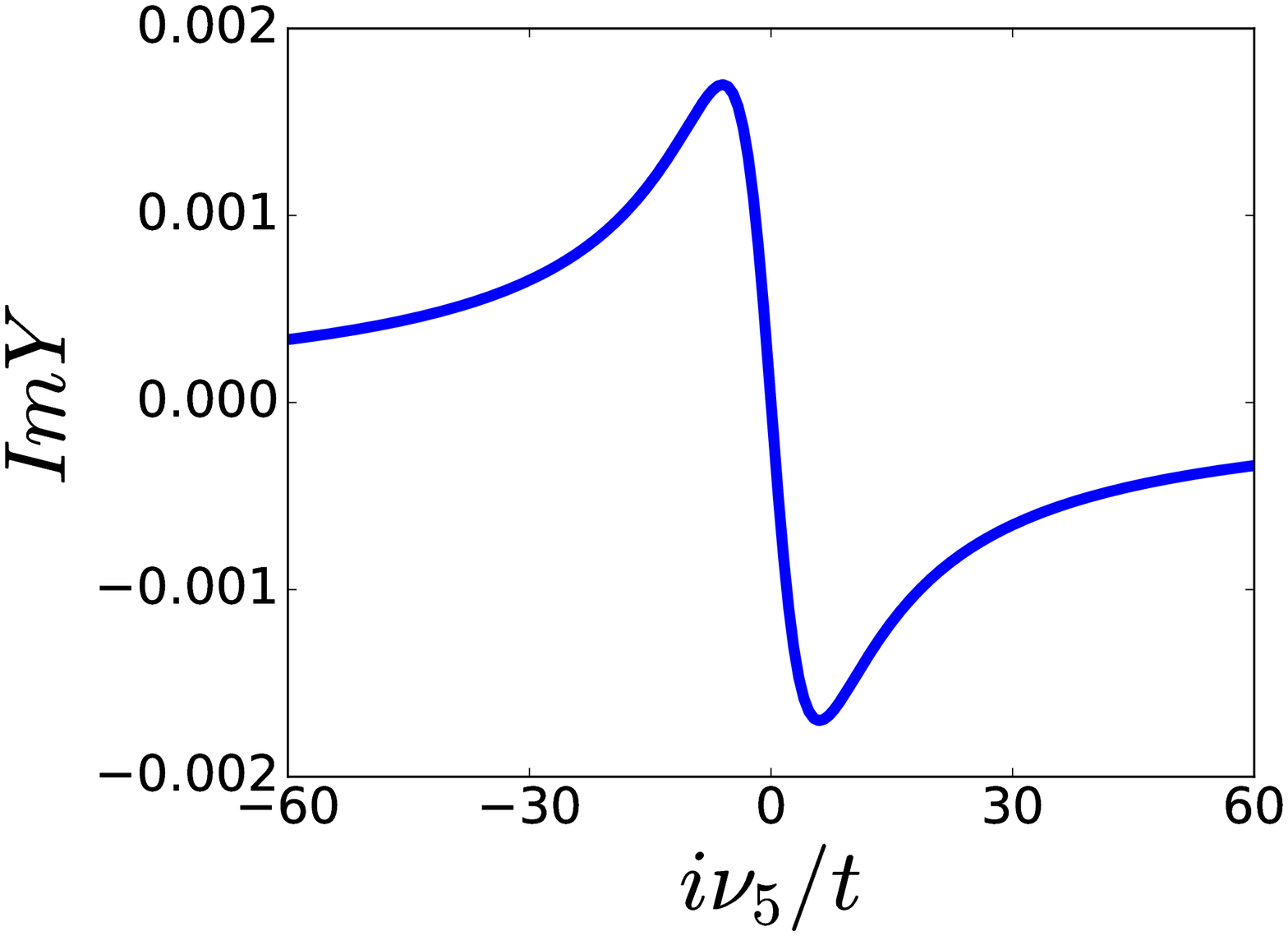}
\includegraphics[width=0.3\linewidth]{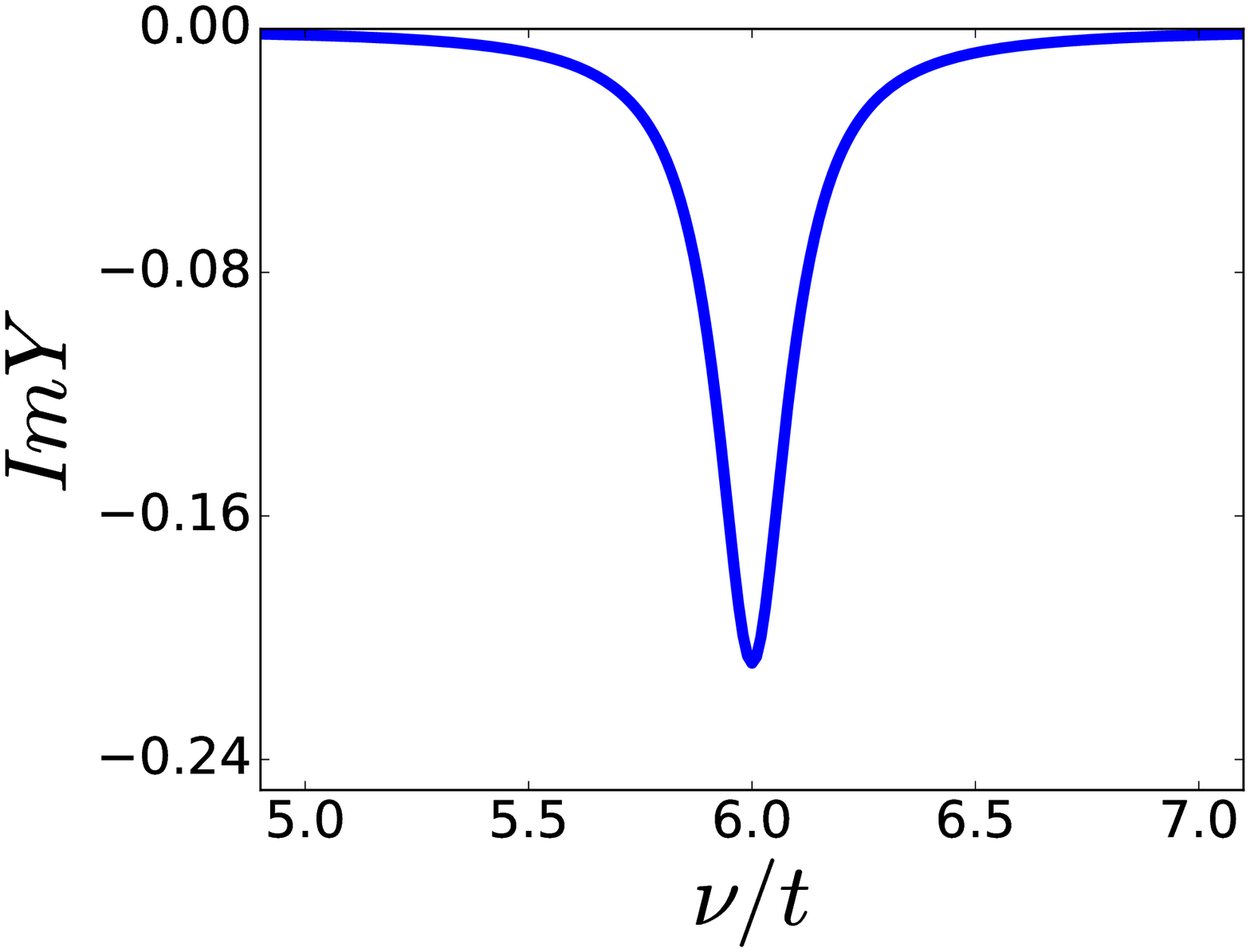}
\caption{\label{fig:combined} \emph{First column}: Feynman diagram; \emph{Second column}: Imaginary part vs. Matsubara frequency; \emph{Third column} Imaginary part vs. real frequency. Data is in unit of $t$ at $\beta=10$, and $\mu=0$ considering a 2D square lattice with lattice constant $a$ for all plots. \emph{Top}: $\Sigma^{(2)}$ at crystal momentum point $\vec k_1=(0,0)$, $\vec k_2 = \vec k_3 = (\frac{\pi}{a}, \frac{\pi}{3a})$; \emph{Middle}: $\Lambda^{env}$ at crystal momentum point $\vec k_1=(0,0)$, $\vec k_2 = \vec k_3 = \vec k_4 = \vec k_5 = \vec k_6 = (\frac{\pi}{a}, \frac{\pi}{3a})$ choosing $\nu_4=\frac{\pi}{\beta}$ and $\nu_6=0$ in Matsubara space and $\nu_4=\nu_6=0$ on real frequency axis; \emph{Bottom}: $Y$ at crystal momentum point $\vec k_1=(0,0)$, $\vec k_2 = \vec k_3 = \vec k_4 = \vec k_5 = (\frac{\pi}{a}, \frac{\pi}{3a})$. We set $\delta=0.1$ in the analytic continuation process $i\nu_n \to \nu + i\delta$. \label{fig:Examples}}
\end{figure}

\emph{Examples: }
To illustrate the utility of AMI we evaluate the temporal integrals of 3 diagrams of increasing complexity shown in the left hand column of Fig.~\ref{fig:Examples}.
 We assume a 2D square lattice with tight-binding  dispersion  $\epsilon_{\vec k} = -2t( \cos (k_xa) + \cos (k_ya)) -\mu$ where $t$ is the hopping amplitude, $a$ is the lattice constant and $\mu=0$ for simplicity.  The three diagrams are: $\Sigma^{(2)}$, a 2nd order self energy diagram with a single external line; $\Lambda^{env}$ a highly connected 4th order irreducible diagram\cite{Rohringer12} with multiple external frequencies and three independent Matsubara frequencies; $Y$ a 4th order example including four independent frequencies.  The diagrams are translated, save for factors of $U^{n_v}$ as
\begin{align}\label{E: Sigma}
&\Sigma ^{(2)} = \frac{1}{\beta^2}\sum_{\nu_1, \nu_2} G(i\nu_1)G(i\nu_2)G(i\nu_3+i\nu_2-i\nu_1),\\
&\label{E: Lambda_Def}\Lambda ^{env} =\frac{1}{\beta^3}\sum_{\nu_1, \nu_2, \nu_3} G(i\nu_1)G(i\nu_2)G(i\nu_3)G(i\omega)G(i\eta)G(i\theta),  \\
&Y =\frac{1}{\beta^4}\sum_{\{\nu_i\}_{i=1}^{4}} G(i\nu_1)G(i\nu_2)G(i\nu_3)G(i\nu_4)G(i\omega)G(i\theta)G(i\eta). \label{eqn:y}
\end{align}  
The AMI algorithm produces symbolic results in the form of $P_p$, $S_p$ and $R_n$ (see Supplementary Information for explicit forms) which are used to evaluate each diagram,
%
%
%
%
%
\begin{align}\label{E: Y_analytic}
\Sigma ^{(2)} &\to (S_1 * f(P_1)) \times (S_2 * f(P_2)) \cdot R_2^{\Sigma} \\
\Lambda ^{env}  &\to (S_1 * f(P_1)) \times (S_2 * f(P_2)) \times (S_3 * f(P_3)) \cdot R_3^{\Lambda} \nonumber \\ \\ 
Y \ \ \  &\to  (S_1 * f(P_1)) \times (S_2 * f(P_2)) \times (S_3 * f(P_3)) \times \nonumber \\ &\ \ \ \ \ (S_4 * f(P_4)) \cdot R_4^{Y} 
\end{align}
There are 4, 32, and 82 terms for $R_2^\Sigma$, $R_3^\Lambda$ and $R_4^Y$ respectively.  These are then evaluated for a choice of internal and external momenta $\{k_n\}$ and external frequencies $\{\nu_\gamma \}$, on either the Matsubara axis or on the real axis via $i\nu_\gamma \to \nu_\gamma +i\delta$ for a choice of small $\delta$.  Results are shown in Fig.~\ref{fig:Examples} on both the Matsubara and real frequency axes for specific choices of $\{k_n\}$ (which would be integrated to evaluate the full diagram).

Computing higher order Feynman graphs using AMI is straightforward.  We provide in the Supplementary Information a particularly complex example for a 9th order diagram where $R_9$ contains 337982 terms assuming simple poles but the number of terms when treated for poles with multiplicity via Eq.~(\ref{E: Res_mul}) grows to the order of $10^9$.  We note that in general the times $t_c$ and $t_e$ both scale linearly with the number of terms, $\zeta = \prod_p r_p$, where $r_p$ is the number of poles with respect to each integration variable. This results in $\zeta$ growing exponentially in the expansion order but its details depend on the detailed pole structure of a given diagram.

\emph{Concluding Remarks:}
Our approach has two main features. First, the result of AMI, once stored, is equivalent to an analytic result, and is therefore evaluated to machine precision. Furthermore, one can impose analytic continuation symbolically and move to real frequency space without any ill-defined numerical procedure. Second, once $S$, $P$ and $R_n$ are constructed the computational expense for generating the analytic function is small, and the total evaluation time reflects primarily the direct evaluation of the analytic function.  
Once constructed and stored, the function can be evaluated for any set of external variables ($\{\nu_\gamma \}$, $\{k_n\}$, $\{k_\gamma \}$, $U$, $\beta$, and $\mu$) without accumulating error, unlike in DiagMC where one would observe a growth in variance for increasing frequency which worsens for increasing $\beta$. 
In this sense, with AMI the temporal parts of the Feynman integral are solved not only exactly (to machine precision), but also with the lowest possible computational expense, \emph{i.e.} the evaluation of the analytic result. 

In our three examples we have evaluated each diagram for a particular set of internal $\{k_n\}$ and external momenta $\{k_\gamma \}$. Generally, the evaluation of the remaining spatial integrals can be performed with continuous k-resolution, as in the case for DiagMC. 
Our results suggest that AMI is able to evaluate diagrams at an order relevant to other state-of-the-art methods while incurring a competitive computational cost.
In addition, the symbolic result of AMI for each diagram, once constructed, can be applied to any diagram with the same topology given the initial set of $\epsilon^j$ dispersions.  This leads to an interesting possibility that each configuration could be systematically evaluated and stored without need to ever reconstruct the $S$, $P$, and $R_n$ arrays.  Once stored, those arrays can be loaded into memory and systematically evaluated for an arbitrary Hubbard-like problem of arbitrary spatial dimension and dispersion.

Finally, we have presented only the most straightforward algorithm but appreciate that optimizations likely exist. These might include improved routines for manipulating and storing the matrices of typically sparse $\alpha_p^j$ vectors, or approximation schemes whereby terms with small contributions are identified and never evaluated. While in this work we applied the method to single-band systems with constant vertices, extension to non-constant vertices or multi-band systems should be explored.\cite{iskakov:2018,iskakov:2016,gukelberger:2017,inchworm}

\section{Acknowledgments}
JPFL would like to thank Phillip E.C. Ashby for fruitful discussions.
This work was supported by the Simons collaboration on the many-electron problem and by the Natural Sciences and Engineering Research Council of Canada (NSERC). Computational resources were provided by Compute Canada via AceNet and Calcul-Quebec. 


\bibliographystyle{apsrev4-1}
\bibliography{refs.bib}

\end{document}